\documentclass[aps,prd,nofootinbib,twocolumn,superscriptaddress,preprintnumbers,balancelastpage,longbibliography]{revtex4-2}

\usepackage{amsmath,amssymb,mathtools,bm}
\usepackage{graphicx, color, hepunits}
\usepackage[dvipsnames]{xcolor}
\usepackage{float}
\usepackage{filecontents}
\usepackage{multirow}
 \usepackage{hyperref} 
\hypersetup{
    colorlinks=true,       
    linkcolor=blue,        
    citecolor=blue,        
    filecolor=magenta,     
    urlcolor=blue          
}
\usepackage[utf8]{inputenc}
\usepackage[english]{babel}
\usepackage{amsmath,amssymb}
\let\vec\mathbf

\newcommand{\Msun}{M_\odot}
\newcommand{\myol}[2][3]{{}\mkern#1mu\overline{\mkern-#1mu#2}}

\begin{document}

\title{Red-Giant Branch Stellar Cores as Macroscopic Dark Matter Detectors }

\author{Christopher Dessert}
\affiliation{Leinweber Center for Theoretical Physics, Department of Physics, University of Michigan, Ann Arbor, MI 48109 U.S.A.}
\affiliation{Berkeley Center for Theoretical Physics, University of California, Berkeley, CA 94720, U.S.A.}
\affiliation{Theoretical Physics Group, Lawrence Berkeley National Laboratory, Berkeley, CA 94720, U.S.A.}

\author{Zachary Johnson}
\affiliation{Leinweber Center for Theoretical Physics, Department of Physics, University of Michigan, Ann Arbor, MI 48109 U.S.A.}

\date{\today}

\begin{abstract}
We show that macroscopic dark matter (DM) impacts on the degenerate helium cores of red-giant branch (RGB) stars can ignite helium fusion via DM-baryon elastic scattering. The onset of helium burning leads to a characteristic drop in luminosity and rise in temperature that marks the transition to a horizontal branch star. We show that such impacts can alter the RGB luminosity function of globular clusters (GCs), focusing in particular on the GC M15.
Using models of M15 stars constructed with the stellar simulation code MESA, we compute the expected DM-ignition event rates and the theoretical RGB luminosity functions under the null and signal hypotheses. We constrain DM with masses $10^{17}\ {\rm g} \lesssim m_{\chi} \lesssim 10^{20}\ \rm{g}$ and geometric cross sections $10^2\ {\rm cm}^2 \lesssim \sigma_{\chi n} \lesssim 10^{7}\ \rm{ cm}^2 $ assuming that the DM in M15 is sourced by the background Milky Way halo. 
We also place more stringent constraints assuming
that M15 formed in a DM subhalo that survives today.
\end{abstract}

\preprint{LCTP-21-30}

\maketitle

\tableofcontents

\section{Introduction}
Dark matter (DM) is known to exist from a multitude of gravitationally-based evidence accumulated over a century of observation~\cite{Bertone:2016nfn}. Much of the unconstrained DM mass range, which spans from $\sim$ $10^{-22}$ eV fuzzy dark matter~\cite{Hu:2000ke} to $\sim$ $5 M_\odot$ MACHOs~\cite{Carr:2020xqk,2014ApJ...790..159M,Brandt:2016aco},
lies above the mass scales involved in the Standard Model (SM). Elementary DM can exist below the Planck scale, however, more massive DM must be a bound state, typically thought of as comprising new particles. In this case the DM sector should have a particle-antiparticle asymmetry to allow bound state coalescence. A wide and continuous range of interactions can generate DM bound states with masses ranging from very large nuclei, $m\lesssim~M_{Pl}$ \cite{Krnjaic:2014xza,Hardy:2014mqa,Gresham:2017cvl}, to nuggets of mass~$10^{-7}~\text{ g}~\lesssim~m~\lesssim~10^{23}$~g~\cite{Bai:2018dxf}, and up to substructure of mass~$\gtrsim~M_\odot$ \cite{Chang:2018bgx,Buckley:2017ttd}.   
Macroscopic DM can also be comprised of SM particles or contain SM charges if the bound state is energetically prevented from decaying to nucleons \cite{Witten:1984rs,Bai:2018vik,Lynn:1989xb}.    

Macroscopic DM candidates have sufficiently low event rates as to render microscopic direct detection experiments, with exposure times $\sim$1 kT yr, ineffective. However, massive, old astrophysical objects can act as DM detectors with exposure times $\sim$~$\Msun\times$Gyr~$\approx~10^{33}$~kT~yr. For example, it is well known that the passage of macroscopic DM through white dwarfs (WD) can deposit sufficient energy in a local region to ignite a Type 1a supernova~\cite{Graham:2018efk}. Similarly, macroscopic DM passage in neutron stars can ignite superbursts~\cite{Sidhu:2019kpd}. In stars, supersonic DM can dissipate energy in the form of shockwaves, which travel to the surface, releasing transient UV radiation~\cite{Das:2021drz}. Macroscopic DM has also been probed through a host of additional mechanisms that do not require such extreme exposure times, albeit at lower masses~\cite{1984Natur.312..734D,1988PhRvD..38.3813P,SinghSidhu:2018oqs,SinghSidhu:2019loh,Sidhu:2019fgg,Starkman:2020sbz,Cooray:2021dvp}.

In this work, we demonstrate that macroscopic DM can ignite helium fusion in red-giant branch (RGB) stellar cores. An RGB star is a low-mass star which has completed hydrogen burning on the main sequence but has not yet begun helium burning. It has an inert electron-degenerate pure-helium core surrounded by a hydrogen-fusing shell. In the absence of macroscopic dark matter, shell fusion will continue to heat the core for $\sim 0.5$ Gyr, over which time  the luminosity of the RGB star rises~\textemdash\  this continues until the critical temperature for helium ignition is reached in the core. Because degeneracy pressure is independent of temperature, the energy released heats the core, further increasing the fusion rate. The entire core ignites in a runaway reaction known as the ``helium flash" (HF). The only outwardly observable signal is the essentially instantaneous drop in luminosity of the star over $\sim$ $10-50$ kyrs, due to expansion and subsequent cooling. The transition of the RGB star onto the next stage of stellar evolution, the horizontal branch (HB), then takes place over $\sim$ $2$ Myrs.

The HF can occur prematurely if significant energy is deposited in the core. When a macroscopic DM particle traverses the degenerate core of an RGB star, elastic collisions between DM and the stellar material can generate enough heat to initiate local helium fusion, in an early-onset HF. The star undergoes the transition to the HB which in the extreme would entirely eliminate the RGB as a phase of stellar evolution. To capture cases in which only some fraction of RGB stars are subject to a DM-induced HF, we constrain DM using the measured RGB luminosity function (LF), which is the observed number of RGB stars as a function of their luminosity.

The RGB core temperatures and densities increase with time, so the high-luminosity RGB stars already closest to undergoing the standard HF are more easily ignited by DM. The presence of macroscopic DM steepens the LF, since fewer high-luminosity stars survive. To probe this new mechanism, we use the MESA code~\cite{Paxton_2010} to simulate the evolution of the RGB stars in the globular cluster (GC) M15 and compute theoretical LFs to compare to the M15 LF computed in Ref.~\cite{2014PASP..126..733F}.  Old GCs such as M15 are good targets for this search because they are well-studied and host many RGB stars which were formed in similar conditions, {\it e.g.} in similar metallicity environments. We focus on M15 because it is the only GC for which an LF has been directly constructed and presented. The LF can be computed for many other GCs using their Hertzsprung-Russell diagrams, but this is beyond the scope of this work. Additionally, as we will discuss, if the GC formation was seeded by an early universe DM overdensity, then the surviving DM density may be orders of magnitude greater than the Milky Way (MW) density at that location. We assume the standard Milky Way DM halo density for our fiducial constraint, but discuss the much stronger limit obtained if future measurements determine that M15 hosts its own DM halo. 

In Sec.~\ref{sec:theory}, we examine the physics of the HF and the DM-induced ignition mechanism. In Sec.~\ref{sec:mesa}, we detail the MESA simulations of the M15 RGB stars used in this work. In Sec.~\ref{sec:rate}, we discuss the computation of the DM-induced HF rate. In Sec.~\ref{sec:data}, we discuss our construction of the LFs and show the 95\% limits on macroscopic DM parameters from the observed M15 LF. Finally, we conclude and discuss additional targets and methods to improve detection prospects in Sec.~\ref{sec:discussion}.

\section{Inducing the HF in Red Giants}
\label{sec:theory}
In order to prematurely ignite the helium core, DM must deposit enough energy to initiate helium fusion before that energy diffuses away. The ignition of significant helium fusion in a large region will generate a stable flame-front, or deflagration~\cite{Timmes:2000abc}. In this work, we assume that such an energy deposition occurs through SM elastic scattering off macroscopic DM with sufficient interaction strength that the resulting cross section is geometric.

For such a DM candidate undergoing elastic scattering, the average energy transfer per nuclei is $\mathcal{O}(m_n v_\chi^2)$, giving a linear energy deposition rate of \cite{DeRujula:1984axn}
\begin{align}
\frac{dE}{dx}  = \sigma_{\chi n} \rho_{\star} v_\chi^2. \label{eqn:dEdx}
\end{align}
Where $\rho_*$ is the stellar density, which varies strongly throughout the star, and $v_\chi$ the DM velocity at that depth in the star. The DM deposits energy in a cylinder with length a substantial portion of the star, and cross-sectional area given by the geometric cross section. This energy deposition has only local effects, and is hence unobservable (but see~\cite{Das:2021drz}), unless it ignites a runaway fusion reaction. A runaway reaction will occur when the local energy generation rate by fusion exceeds the energy loss rate, which is dominated by diffusion. To estimate these rates, we must assume a temperature profile immediately after the DM impact. We work in cylindrical coordinates where the longitudinal axis points along the direction of the DM path and $r$ is the distance from the center of the DM path. We assume the energy deposition is linearly increasing from the edge of the cylinder ($r=\sqrt{\sigma_{\chi n}/\pi}$) to the center, yielding
\begin{align}
    \frac{dE}{dV}=\frac{3}{\sigma_{\chi n}}\left(1-\dfrac{r}{\sqrt{\sigma_{\chi n}/\pi}}\right) \frac{dE}{dx}.\label{eqn:dEdV}
\end{align} 
This profile is just one example of a heat deposition profile. We vary the energy deposition profile and find that if they are not sharply peaked at the center, the effect on the ignition parameter can be $\mathcal{O}(1)$; see Appendix \ref{app:ignition} for further discussion.
We then compute the temperature $T_{\rm hot}$ that the energy deposition heats the stellar material to, via numerically solving 
\begin{align}
    \frac{dE}{dx} =\rho \sigma_{\chi n} \int^{T_{\rm hot}}_{T_*} c_v dT.
\end{align}
We use analytic approximations for the heat capacities of the electrons, ions and photons, which are
\begin{align}
c_v^{e^-} &=
\left\{
\begin{array}{ll}
      \dfrac{7 \pi^2}{15} \dfrac{k_B^4}{\hbar^3 c^3} \dfrac{T^3}{\rho_*} & T> 10^9 K \vspace{0.2cm}\\
      \dfrac{3 k_B}{m_{^4 {\rm He}}}   & T_{degen} < T< 10^9 K \vspace{0.2cm}\\
      \dfrac{4 k_B^2 T}{m_{^4 {\rm He}} E_F \pi^2}  & T<T_{degen},
\end{array}\right. \\
c_v^{\gamma} &= \frac{4}{15} \pi^2\frac{k_B^4 }{\hbar^3 c^3} \frac{T^3}{\rho_*}, \\
c_v^{ion} &= \dfrac{3 k_B}{2m_{^4 {\rm He}}}. \label{eqn:cv electron}
\end{align}
where~$T_{degen}=~3 E_F/\pi^2 k_B$, for $E_F$ the Fermi energy, and $10^9$ K approximates the transition to a relativistic electron gas. We neglect the relativistic degenerate electron regime because the electron contributes negligibly to the heat capacity when it applies.

The dominant fusion process is helium fusion, specifically through the triple-alpha process, wherein three $^{4}$He fuse through the resonant Hoyle state to create a $^{12}$C nucleus. The triple-alpha energy generation rate is~\cite{Timmes:2000abc}
\begin{align}
\dot{S}_{3\alpha} = 5.1 \left(\frac{\rho_*(r)}{10^4 {\rm\ g/cm^2}}\right)^2 \left( \frac{X({\rm He})}{T_9(r)}  \right)^3 e^{-4.4/T_9(r)} {\rm \frac{ergs}{s\ g}},\label{eqn:S3rate}
\end{align}
where $T_9(r)\equiv T(r)/(10^9$ K). Electron screening enhances the rate at low temperatures, however this effect is $\lesssim 5\%$ at $5 \times 10^8$ K, where the nuclear generation rate is already quite low and so is neglected. Note that carbon fusion, which operates at similar temperatures, is suppressed in the initial ignition by the low carbon abundance.
Thus we use Eq.~\eqref{eqn:S3rate} to determine the local energy generation rate, which can be expressed as
\begin{align}
\frac{d\dot{E}_{\rm nuc}}{dx}&= 2 \pi \int \dot{S}(r) \rho_* r dr.\label{eqn:NuclearRate}
\end{align}
The heated region cools off via diffusion at a rate of 
\begin{align}
\frac{d\dot{E}_{\rm diff}}{dx} = - 2 \pi k r \frac{dT}{dr}\bigg |_{r=\sqrt{\sigma/\pi}}\label{eqn:Diffusion}
\end{align}
where $k$ is the thermal conductivity. Runaway nuclear fusion will occur if the generated heat is greater than that lost through diffusion. To see this explicitly, we define an ignition parameter, $\zeta$, as 
\begin{align}
     \zeta &\equiv \frac{d\dot{E}_{\textrm{nuc}}/dx}{d\dot{E}_{\textrm{diff}}/dx}.\label{eqn:IgnitionRatio}
\end{align}
Ignition occurs for $\zeta>1$. We will see that for typical points in our constrained region, this ratio is far greater than $1$.

\section{MESA Simulations of M15 RGB stars}
\label{sec:mesa}

   \begin{figure*}[!t]
    \includegraphics[width = 0.99\textwidth]{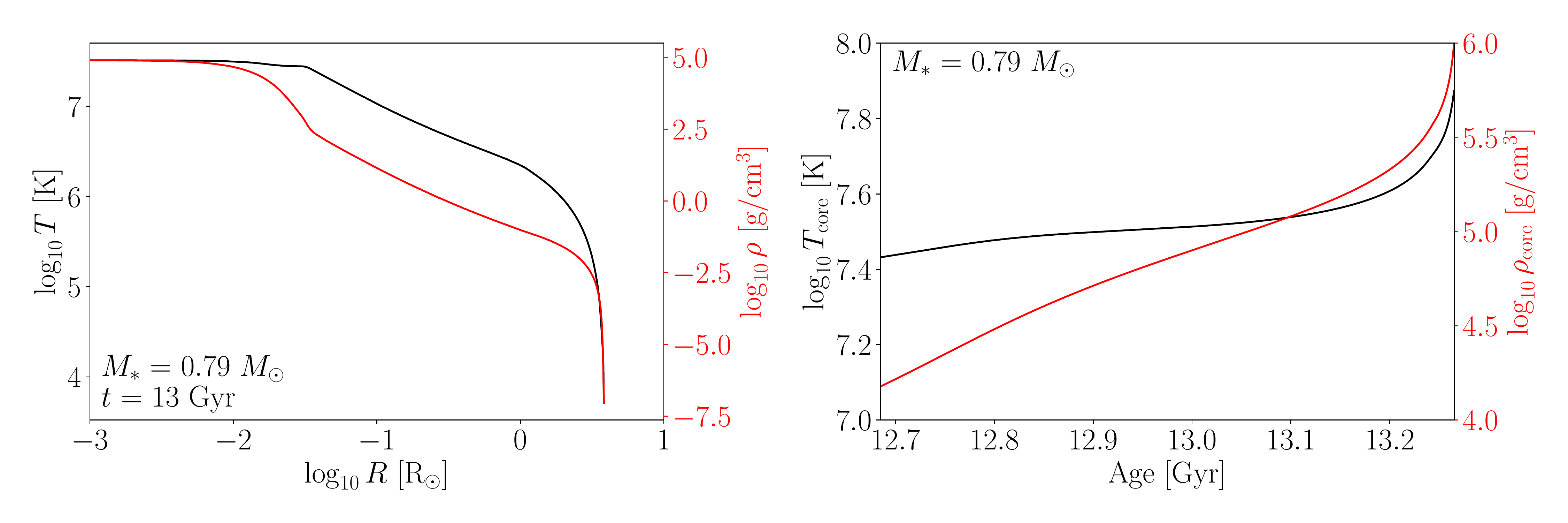}
\caption{ \textbf{Left Panel} The internal temperature (black) and density (red) profiles of the 0.79 $\Msun$ star when it is 13 Gyr old. \textbf{Right Panel}  The central temperature (black) and central density (red) of the 0.79 $\Msun$ star as a function of time during the RGB phase, which spans the x-axis of the plot.}
\label{fig:Trho}
\end{figure*}

   \begin{figure}[!t]
    \includegraphics[width = 0.49\textwidth]{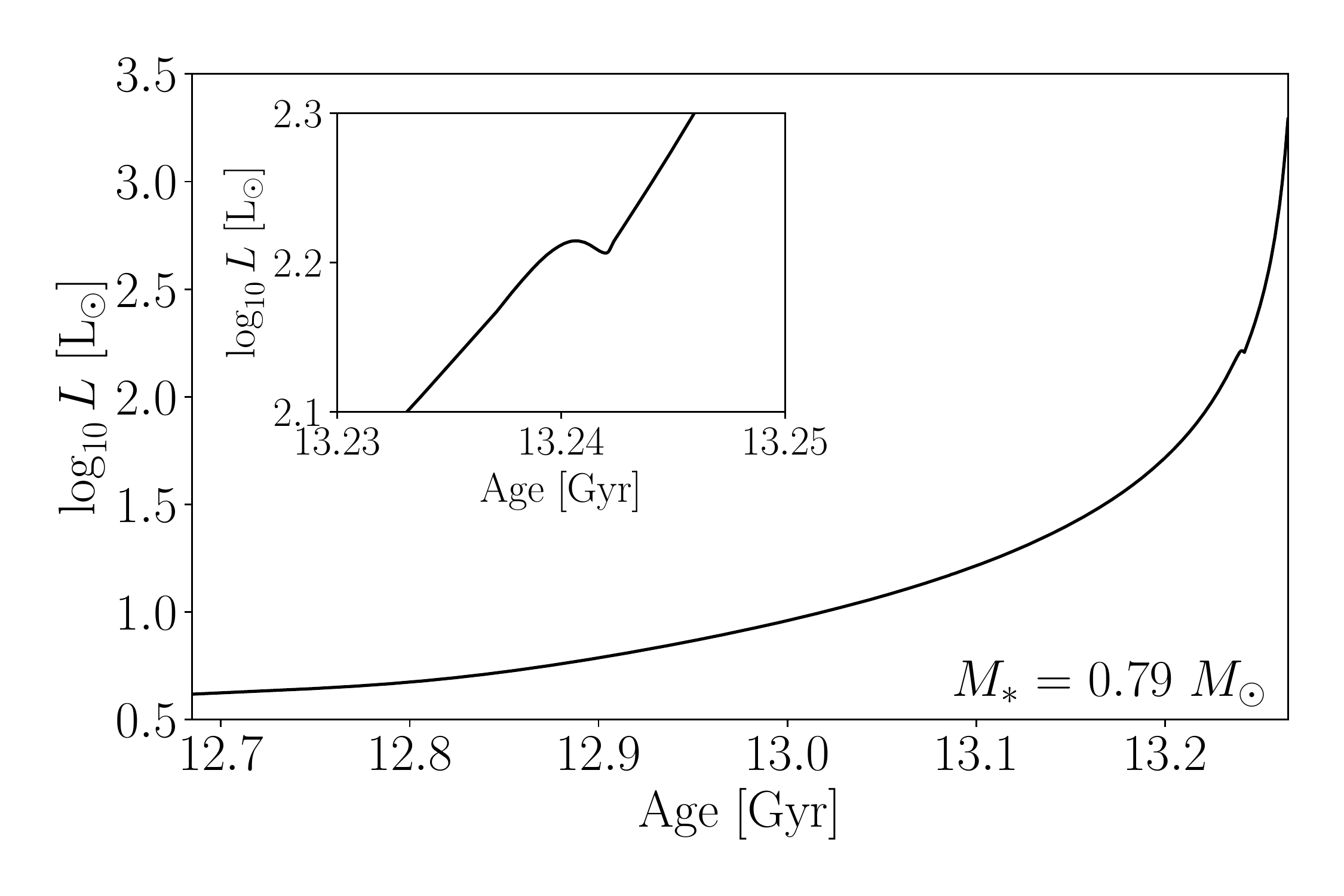}
\caption{\textbf{Main Plot} The luminosity of the 0.79 $\Msun$ star as a function of time during the RGB phase, which spans the x-axis of the plot. \textbf{Inset} The same as the main plot but zoomed in around the red giant bump.}
\label{fig:Lum}
\end{figure}

In this section, we describe our MESA simulations of the RGB stars in the GC M15. We use MESA version 12115~\cite{Paxton_2010} to simulate the evolution of the RGB stars. The age of M15 is $13 \pm 1$ Gyr~\cite{2014PASP..126..733F}; we fix its age at 13 Gyr. Under this assumption, the masses of the RGB stars in M15 range from $0.785 \Msun$ (just ascending the RGB) to $0.795 \Msun$ (at the end of the RGB phase). The metallicity of the M15 GC is [Fe/H]$~=~-2.1$~\cite{2014PASP..126..733F}, which corresponds to an average isotopic abundance of [Z]$~=~-1.57$~\cite{2013ApJ...774...75W}~\textemdash~we fix this to be the initial abundance in the MESA simulations.

We simulate stars of initial mass $M_*$ between $0.785$ and $0.795 \Msun$ in steps of $0.001 \Msun$  from the pre-main sequence through the beginning of central helium ignition. The MESA simulations return one-dimensional (radial) profiles of temperature $T$, density $\rho$, and isotopic compositions at many time points throughout the evolution, along with bulk properties of the star: age $t$, luminosity $L(t)$, effective temperature $T_{\rm eff}(t)$ and surface gravity $g(t)$. We use these stellar models to (i) compute the expected DM-induced HF rate in each star and (ii) compute the theoretical LFs under the null and signal hypotheses.

In the left panel of Fig.~\ref{fig:Trho}, we show the temperature and density profiles, at $t = 13$ Gyr, of the star with mass $\myol{M}$ in the center of the range, $\myol{M}~\equiv~0.790~\Msun$. Over the full mass range we focus on in this work, the stellar properties in the RGB phase are the same as those shown here to $< 1\%$. The helium core occupies $\sim$1$\%$ of the stellar radius, when the temperature and density sharply drop traversing farther out of the star. The vast majority of the star by volume is the low-density stellar envelope, which at this time extends out to $\sim$4 $R_\odot$, though the envelope will continue to expand in size until the HF. In the right panel, we show the time-evolution of the central temperature and density, which sharply increase as the star nears the flash. The lower and upper x-limit of the plot corresponds to when the star enters the RGB phase at $\sim$12.68 Gyr and when the HF occurs at $\sim$13.27 Gyr, corresponding to a total RGB duration for this star of $T_{\rm RGB}^{\myol{M}} = 590$ Myr. We parameterize the fraction of the RGB branch ascended by the star by $\tau$, 
\begin{equation}
    \tau\equiv\dfrac{t-t_{\rm TAMS}}{T_{\rm RGB}},
\end{equation}
where $t_{\rm TAMS}$ is the terminal age main sequence, the age at which the star begins to ascend the RGB, and $T_{\rm RGB}$ is the length of the RGB phase. Therefore $\tau=0$ defines when the star begins the RGB phase and $\tau=1$ when the standard HF occurs and the star moves onto the HB. Note that the explicit mapping between $t$ and $\tau$ depends on the stellar mass through $t_{\rm TAMS}$ and $T_{\rm RGB}$.

In Fig.~\ref{fig:Lum}, we show the luminosity evolution of the star. Based on this plot, we can reconstruct the two basic features of the LF by making the approximation that there are the same number of stars in each age interval. Firstly, because the luminosity increases faster with increasing age, there will be more stars at lower luminosities. Therefore, the LF will sharply increase towards lower luminosities. Secondly, at $\sim$13.24 Gyr, the luminosity abruptly stops increasing for a few Myr (see the inset). As the hydrogen-burning shell expands outwards, it reaches a composition discontinuity created by the first dredge-up, which leads to a brief halt in the luminosity increase. Because of this, stars spend a longer time at this luminosity level. This is known as the red giant bump~\cite{1967ZA.....67..420T,1968Natur.220..143I}, and indeed in the LF there is a small bump at this luminosity.

\section{The DM-Induced He-ignition Rate}
\label{sec:rate}

\begin{figure*}[!t]
    \centering
    \hspace*{-0.025in}\includegraphics[width = 0.993\textwidth]{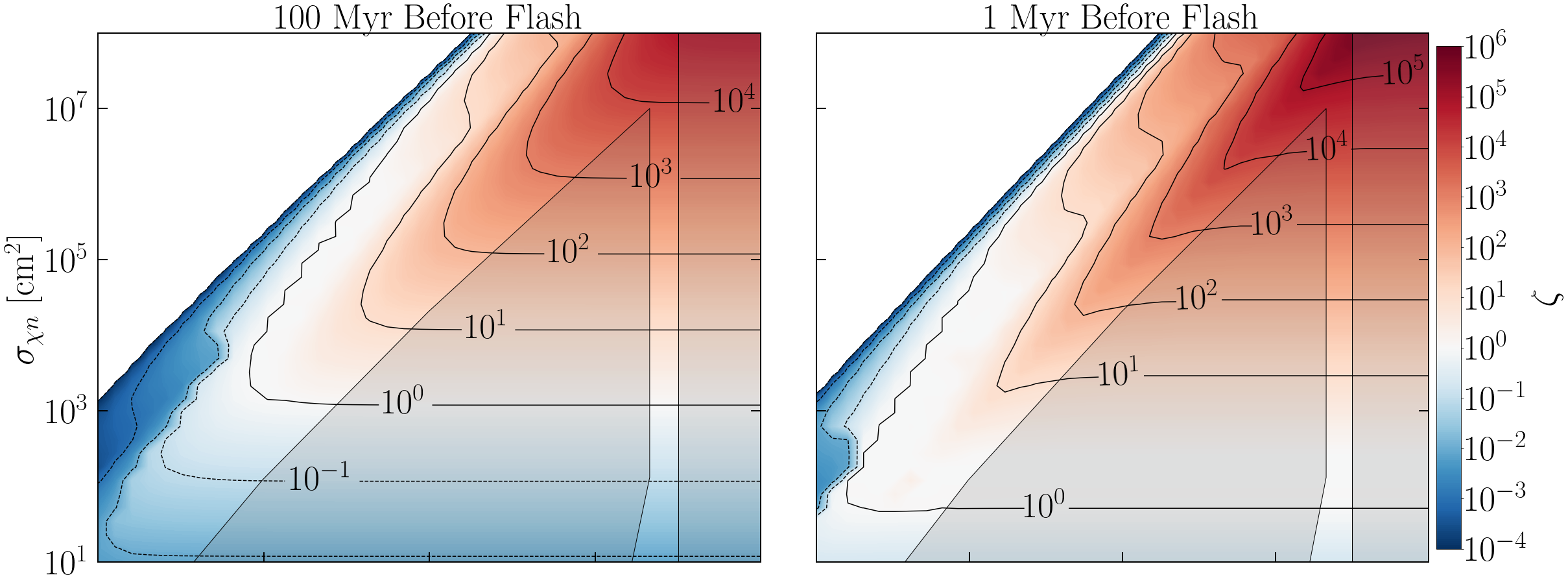}\vspace{-0.07in}
    \includegraphics[width = \textwidth]{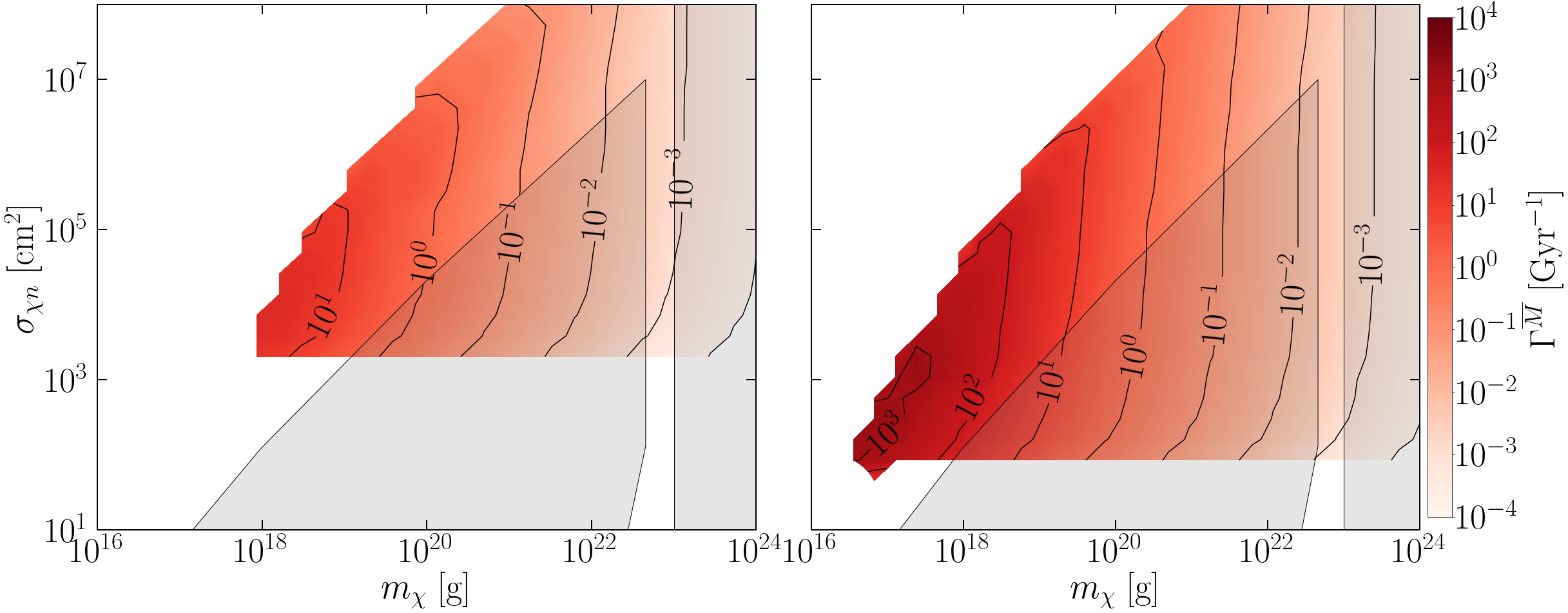}
\caption{ \textbf{Top Panels:} The maximum ignition parameter, $\zeta$, described in Eq.\eqref{eqn:IgnitionRatio}, for a simulated DM trajectory assuming an initial angular momentum  $\ell_0 = 10^{-4} c \times 10^{10}$ cm, where $10^{10} \text{cm} = 0.14\ R_{\odot}$, and the benchmark mass $\myol{M}$ star. In light grey are constraints from non-observation of DM-induced Type Ia supernova in WDs~\cite{Janish:2019nkk,Sidhu:2019kpd}(triangular shape) and from microlensing~\cite{Smyth:2019whb} (vertical line). \textbf{Top Left:}  100 Myr before the expected HF ( total RGB phase duration is $T_{\rm RGB}\sim 590$ Myr).    \textbf{Top Right:}  1 Myr remaining in the RGB phase. \textbf{Bottom Panels:} The rate of DM induced HF ignition $\Gamma^{\overline{M}} $[Gyr$^{-1}$], with correction factor $c(t)$ set to $1$ for comparison purposes, see Eq.~\eqref{eqn:MWrate}. \textbf{Bottom Left:}  $100$ Myr before the nominal HF. \textbf{Bottom Right:} Ignition rate $1$ Myr before HF.}
\label{fig:Ignition Ratios}
\end{figure*}

The rate at which DM trajectories ignite a given stellar core depends sensitively on the details of the DM distribution. There are two basic assumptions one can make; either the GC formed within a DM overdensity that survives today, or that it did not and the only DM around is that of the Milky Way (MW) halo. This first assumption is nominally the expectation of the standard model of cosmology, and would give a DM density orders of magnitude above the MW halo density.
However, there is a lack of supporting observational evidence. In \cite{Ibata:2013} a prototypical GC was found to contain very little DM under modest assumptions, and
\cite{2010MNRAS.405..375G} found that GCs closer than $100$ kpc to the Galactic Center typically have halos that have already merged with the galaxy halo (M15 is $\sim 10$ kpc from the Galactic Center). 
Numerical simulations of the GC DM halo scenario find that tidal stripping can significantly reduce the amount of DM, however this still would likely leave some portion of the DM within the tidal radius \cite{Saitoh:2005tt,2010MNRAS.405..375G,Creasey1806,Mashchenko:2004hj}, and this scenario is often assumed in the literature \cite{HESS:2011yps,Leane:2021ihh,Das:2021drz}. 

Due to the lack of a complete consensus, we place our fiducial limits  assuming the DM density in M15 is purely that of the MW DM halo, and defer our analysis including a GC DM halo to App.~\ref{app:GChalo}. 

We model the MW halo with the NFW profile~\cite{Navarro:1995iw,Navarro:1996gj} from~\cite{Evans:2018bqy} with scale factor $r_s = 20$ kpc, normalized such that distance of Earth to the Galactic Center is $8.12$ kpc~\cite{GRAVITY:2018ofz}. In the context of this profile, we infer a MW halo DM density at the location of M15 today, 10.76 kpc from the Galactic Center~\cite{2019MNRAS.482.5138B}, of $\rho_{\text{DM}}~=~0.35~\pm~0.10$~GeV~cm$^{-3}$, and we assume the lower 1$\sigma$ value $0.25$~GeV~cm$^{-3}$ in our analysis. However, M15 has a galactic orbital period of $\sim$140 Myr with a $\sim 3.7-4$ kpc periapsis, so that its local DM density periodically changes by a factor $\sim$4.8 over the length of the RGB phase, $\sim$600 Myr. This orbit was computed numerically in Ref.~\cite{Baumgardt2019} by evolving the M15 orbital trajectory backwards in the galactic gravitational potential in \cite{Irrgang2012}, and in our analysis we assume the best-fit orbit provided. We assume the DM velocity distribution is given by the standard halo model,
\begin{align}
\begin{split}
    f({\bf v},t) =& N \text{exp}\left(-\frac{({\bf v + v_{GC}}(t))^2}{2 \sigma_{vr}^2}\right) \\ 
    & \times \Theta(v_{esc}-|\vec{v}+\vec{v}_{GC}(t)|).
    \label{eqn:SHM}
\end{split}
\end{align}
for normalization $N$, radial velocity dispersion given by the upper $1\sigma$ value from~\cite{Evans:2018bqy}, $\sigma_{vr} = 167$ km s$^{-1}$, and a MW escape velocity $v_{esc}$ of $544$ km s$^{-1}$. The stellar velocity within M15 is $\approx 10 $ km s$^{-1}$, thus to a good approximation all stars move at the GC velocity. At the present location of M15, we find $v_{GC}=123~\pm~1.1\ $km~s$^{-1}$.

The depth to which the DM must penetrate is to the ignition radius of the star $R_{\zeta=1}(\tau)$, where again $\tau$ is the proportion of the RGB branch a given star in M15 has gone through. For DM with initial velocity $v_0$, the rate of impacts within radius $R_{\zeta=1}$ is enhanced by a gravitational focusing factor of $\left(1+ \left(v^{*}_{\text{esc}}(r)/v_0 \right)^2 \right)$, where $r$ is the distance of the DM from the center of the star. The $v_{\text{esc}}$ at the core is $\sim 10^3$ km $s^{-1}$, so we approximate this effect with $\left( v_{*\text{esc}}(r)/v_0 \right)^2$.\footnote{One may also think there is a gravitational focus from the GC potential, however this effect is not independent of the stellar focusing and including this effect increases the rate only by $v^2_{GC\text{esc}}/v^2_{*\text{esc}}\sim 10^{-4}$. } These effects can be analytically integrated, which we show in App.~\ref{app:rates}. 
We divide these contributions to the DM-induced ignition rate into contributions that depend only on stellar evolution ({\it i.e.}, the time-dependence is given only by $\tau$) and those that depend on the GC location throughout the galaxy, where the time-dependence is on $t$ directly. 
\begin{align}
\Gamma(t,\tau) &= \Gamma_0  \times c(t,v_{GC},v_{\text{esc}}) \label{eqn:MWrate} \\
\Gamma_0(\tau) &= 2 \sqrt{2 \pi} \frac{\rho_{DM}(t_0)}{m_{\chi}} R^2_{\zeta=1}(\tau) \sigma_{vr} \times  \frac{ v^2_{*\text{esc}}(R_{\zeta=1}(\tau)) }{ 2\sigma_{vr}^2}
\end{align}
Where $t_0$ implies evaluation at the position of M15 today. The correction factor $c(t,v_{GC},v_{\text{esc}})$ includes the finite MW escape velocity, and the variation of $v_{GC}$ and $\rho_{DM}$ over time as the GC orbits the MW. It varies within $c(t)\ \epsilon\  [0.87,2.47]$ with large values occuring when the GC passes near the galactic center and its higher DM density. We detail its computation in App.~\ref{app:rates} and show our results in Fig.~\ref{fig:correction}.

The value of $R_{\zeta=1}$ is determined by simulation of a DM infall to the benchmark star $\overline{M}=0.79\ \Msun$. We use MESA generated profiles of the stellar density as a function of radius, and the friction force derived from Eq.~\eqref{eqn:dEdx}. We simulate DM trajectories starting at infinity with varying impact parameters $b$, scanning over signal parameters ${\bm \theta_s}$ and at multiple stellar ages $t$. For each point in the scan, $R_{\zeta=1}$ is defined as the largest radius at which the trajectory ignites the star, typically of order $10^9$ cm. We implement Fehlberg's $4^{th}(5^{th})$ order method and use an error tolerance per step of $10$ m and $10^{-9} c$. We verified this procedure by checking that in selected cases the global error estimate was within $10\%$. 
 
An important simplification is that the DM trajectory depends only on the initial angular momentum per unit mass, $\ell=b v_0^{\text{max}}$, rather than individually on $b$, and velocity at infinity, $v_0$, within $\sim 10 \%$ below $v_0\approx 5 \times 10^{-3}c$. Further, the rate of stellar ignition depends only on the maximum angular momentum, $\ell^{\text{max}}$, defined by the largest $\ell$ that satisfies our ignition condition, $\zeta>1$. Trajectories directly towards the star, defined by small $\ell_0$, impact both deeper in the core and at higher velocities, leading to larger $\zeta$, see Eq.~(\ref{eqn:IgnitionRatio}). Trajectories with larger initial angular momentum may result in DM which only skims the core, or in extreme cases undergoes multiple orbits, resulting in lower $\zeta$. Thus all smaller $\ell_0$ will also ignite. This allows us to eliminate a dimension from our parameter scan. Thus explicitly what we determine from simulation is the maximum $\ell=R_{\zeta=1} v_{*\text{esc}}(R_{\zeta=1})$ that causes ignition, yielding a rate of
 \begin{align}
    \Gamma^{\myol{M}}({\bm \theta_s},t) = \sqrt{2 \pi} \frac{\rho_{DM}(t_0)}{m_\chi} \frac{\ell_{\rm max}({\bm \theta_s},\tau)^2}{\sigma_{v_r}} \times c(t,v_{GC},v_{\text{esc}})\label{eqn:Gamma-simple}
\end{align}
for our benchmark star. 

In the top panels of Fig.~\ref{fig:Ignition Ratios} we simulate DM trajectories for fixed initial angular momentum, plotting the maximum ignition ratio $\zeta$ for each $m_\chi$ and $\sigma_{\chi n}$. The left and right panels are for two stars of the same mass $\overline M$ but of different ages, corresponding to 100 and 1 Myr before the standard HF respectively. The triangular shape of each $\zeta$ contour can be understood simply. The series of ``hypotenuses" mark where the DM loses significant kinetic energy in the stellar envelope. The white region begins where the DM loses all initial kinetic energy before reaching the core. The slope of these lines are close to unity as the acceleration from friction is $\propto \sigma_{\chi n}/m_\chi$. Moving perpendicularly away from the hypotenuse towards larger $m_\chi$ and smaller $\sigma_{\chi n}$ increases the depth at which the DM penetrates. Far away from this line, friction is irrelevant so the only relevant parameter for $\zeta$ is $\sigma_{\chi n}$, where lower cross section corresponds to decreased energy deposition by Eq.~\eqref{eqn:dEdx}, until ignition is no longer possible. We can see from the relative location of the $\zeta=1$ contour in the left and right panels that as the star approaches the standard HF, smaller cross sections may ignite the star, consistent with Eq.~(\ref{eqn:dEdV}).

In the bottom panels we plot the simulated rate of igniting DM impacts on a star in M15 for the same two ages. The minimum igniting cross section decreases over time, following the $\zeta=1$ contour in the top panels. Near the minimum $\sigma_{\chi n}$ the contours turn sharply due to the rapid decrease in the portion of the core with sufficient temperature and density to be ignited by a given cross section.
Where the rate is non-zero in both left and right panels, we see the overall rate increases with time due to the larger maximum igniting impact parameter for stars closer to their nominal HF. 

\section{Constraints on Macroscopic DM}
\label{sec:data}

In this section, we outline our modeling of the theoretical LFs under the dark matter hypothesis and assess their goodness-of-fit to the measured M15 LF.

\begin{figure}[!t]
\includegraphics[width = 0.49\textwidth]{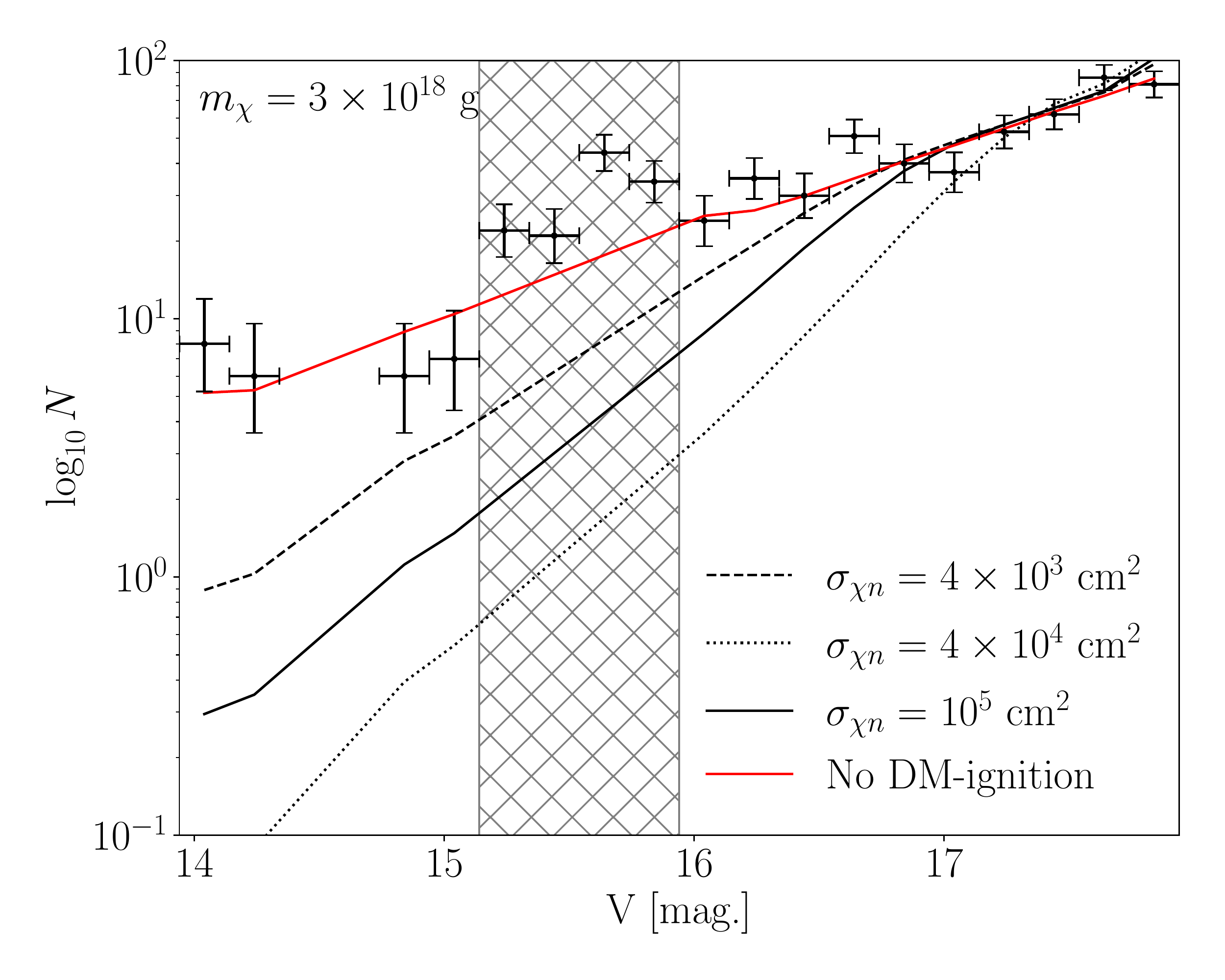}
\caption{ The M15 LF (black data points) with uncertainties given by $\sqrt{N_i}$. In red, we show the best-fit LF with no DM ignition. In black, we show the LFs with $m_\chi = 3 \times 10^{18}$ g and $\sigma_{\chi n} = 4\times10^3$ cm$^2$ (dashed), $\sigma_{\chi n} = 4\times10^4$ cm$^2$ (dotted), $\sigma_{\chi n} = 10^5$ cm$^2$ (solid). The hatched region indicates those four bins around the RGB bump which are masked in the analysis.}
\label{fig:LF}
\end{figure}

Although previously in Sec.~\ref{sec:mesa} we had qualitatively discussed the LF as a function of $L$, the observable is the apparent $V$-magnitude. M15 was observed three times between 2002 and 2004 by the Hiltner 2.4m telescope at the Michigan-Dartmouth-MIT Observatory at Kitt Peak, Arizona. Ref.~\cite{2014PASP..126..733F} used that data to construct the LF between apparent visual magnitudes $V~\in~[13.94,17.94]$ (except $V$ between  $14.34$ and $14.74$) by~\cite{2014PASP..126..733F}, shown in Fig.~\ref{fig:LF}, and we fit our theoretical LFs to that data in this work. Explicitly, we start with the number of stars $N_i$ in bin $i$ corresponding to apparent visual magnitude $V_i$ with bin widths of $0.2$ magnitudes. We will refer to this data as ${\bm d} = \{N_i\}$. In particular, the red-giant bump is visible around 15.5 $V$-magnitudes. We mask the four data points between $15.14$ and $15.94$ $V$, where the model in~\cite{2014PASP..126..733F} did not accurately capture all of the physics of the red-giant bump (our null model, also in Fig.~\ref{fig:LF}, is nearly identical to that found in~\cite{2014PASP..126..733F}). The relation between $L$ and $V$ is given by 
\begin{align}
    V(t;M_*) =& -2.5\log\left(\dfrac{L(t;M_*)}{L_\odot}\right) \\
    & + 4.74 + \mu - BC_{\rm V}(t,T_{\rm eff},g) \label{eqn:LtoV}
\end{align}
where $\mu$ is the dereddened distance modulus with best fit $\hat{\mu} = 14.86$ \cite{2014PASP..126..733F} and we compute bolometric corrections in the $V$-band $BC_{\rm V}$ with the python package \texttt{isochrones}~\cite{2015ascl.soft03010M}.

To construct a theoretical LF for the cluster, we need to know the DM-induced flash rate and the $V$ as a function of time for the range of possible stellar initial masses in the cluster. The former we have detailed the calculation of in Sec.~\ref{sec:rate} for the $\myol{M}$ model.
 To compute the latter, we start from the luminosity-age relations $L(t;M_*)$ generated by MESA and compute $V(t;M_*)$ as in Eq.~\eqref{eqn:LtoV}.

We now make the assumption that the GC evolution is coeval~\cite{2014PASP..126..733F,Tarumi:2021fni}. Then the stellar ages are all $t_{\rm M15} = 13$ Gyr, with V magnitudes $V(t_{\rm M15};M_*)$.
Finally, we invert this relation to obtain the initial stellar mass as a function of $V$ magnitude today, $M_*(V)$. For a particular signal model ${\bm \theta_s}$, the expected number of stars in $V$ magnitude bin $i$ spanning magnitudes $V_i^{\rm min}$ to $V_i^{\rm max}$ is

\begin{align}
    \lambda_i({\bm \theta_s}) = A \times C_i \int_{{M_*(V_i^{\rm min}})}^{M_*(V_i^{\rm max})} dM \dfrac{dN}{dM}(M) P_{\rm surv}({\bm \theta_s},M) \label{eq:LF}
\end{align}
where $A$ is an arbitrary normalization constant to be fit to the data and $C_i$ is the completeness factor in bin $i$ computed via artificial star tests in~\cite{2014PASP..126..733F}. $dN(M)/dM$ is the initial mass function, which we take to have a Salpeter form $dN(M)/dM \propto M^{-2.35}$~\cite{1955ApJ...121..161S}. The integration limits are entirely determined by stellar physics, while the DM physics is only captured in the survival probability $P_{\rm surv}({\bm \theta_s},M)$, which we compute as follows. As detailed in Sec.~\ref{sec:rate}, we have simulated the DM-induced ignition rate $\Gamma^{\myol{M}}({\bm \theta_s},t,\tau)$ for the mass $\myol{M}$ star. Then 

\begin{align}
    P_{\rm surv}({\bm \theta_s},M_*) = \exp\left(-\int_0^{t_{\rm M15}}dt\ \Gamma^{M_*}({\bm \theta_s},t,\tau)\right),\label{eq:Psurv0790}
\end{align}
which implicitly assumes that the DM-induced ignition rates are equal across stellar masses at fixed percentage of time completed on the RGB stage $\tau$. Note that the rates are always zero unless the star is on the RGB.

In Fig.~\ref{fig:LF} we show the theoretical LFs $\lambda = \{\lambda_i({\bm \theta_s})\}$ for three separate cases, along with the measured completeness-corrected LF of M15. The LF with no DM-ignition has the shallowest slope, and the red giant bump is clearly visible in the bin around $V = 15.6$ magnitudes. We then show three curves under the model assumption $m_\chi = 3 \times 10^{18}$ g, at $\sigma_{\chi n} = \{4\times 10^3$ cm$^2$, $4\times 10^4$ cm$^2$, $10^5$ cm$^2$\}. Tracing these curves from high to low $V$, they tend to follow the no DM-ignition line at until a particular $V$ at which DM-ignition turns on, which sharpens the slope of the LF. Recall that stars with lower $V$ are more highly evolved, and so have smaller survival probabilities. The DM model with the smallest cross section $\sigma_{\chi n} = 4\times 10^3$ cm$^2$ is in the region where increasing cross section decreases the survival probability, because the increased cross section increases the energy deposition in the core. The model with the largest cross section $\sigma_{\chi n} = 10^5$ cm$^2$ lies in the region where increasing cross section increases the survival probability because the DM has lost significant kinetic energy in the stellar envelope. The model with middling cross section $\sigma_{\chi n} = 4\times 10^4$ cm$^2$ delineates the boundary between these two regions, and as such has the minimum survival probability for any DM with that mass. The LFs for other DM masses display similar behavior.

With the model in the same form as the data, we can write down the joint likelihood over $V$ magnitude bins

\begin{align}
    \mathcal{L}({\bm d}|{\bm \theta_s},{\bm \theta_{\rm nuis}}) = \prod_i \mathcal{N}(\lambda_i({\bm \theta_s}) - N_i, \sqrt{N_i})\label{eq:likelihood}
\end{align}
where the nuisance parameters ${\bm \theta_{\rm nuis}} = \{A,\mu\}$. $A$, as mentioned previously, is a normalization parameter that sets the total number of RGB stars in M15. $\mu$ is the distance modulus for M15, which we let float around its central value $\hat{\mu}$. 
$\mathcal{N}(x,\sigma_x)$ is the Gaussian likelihood with a mean $x$ and standard deviation $\sigma_x$.

We construct the profile likelihood at fixed mass $m_\chi$ as a function of the cross section $\sigma_{\chi n}$, given by

\begin{align}
    \mathcal{L}({\bm d}|{\bm \theta_s}) = \mathcal{L}({\bm d}|{\bm \theta_s},\hat{{\bm\theta}}_{\rm nuis})\label{eq:profilelikelihood}
\end{align}
where $\hat{{\bm\theta}}_{\rm nuis}$ denotes the values of ${\bm \theta_{\rm nuis}}$ that maximize the likelihood at that ${\bm \theta_s}$. To define the region ruled out by this analysis, we construct the test statistic~\cite{Cowan:2010js}

\begin{align}
    q({\bm \theta_s}) =
 \left\{ 
\begin{array}{ll}
2 \left[ \ln \mathcal{L}({\bm d} \vert m_\chi, \hat \sigma_{\chi n}) - \ln \mathcal{L}({\bm d} \vert {\bm \theta_s}) \right] &  \hat \sigma_{\chi n} \leq \sigma_{\chi n} \vspace{0.05cm}\\
 0 & \hat \sigma_{\chi n} > \sigma_{\chi n}
 \end{array} \right.\label{eq:TS}
 \end{align}
where $\hat \sigma_{\chi n}$ is the value of the cross section that maximize the likelihood at that mass. We exclude at 95\% confidence the cross sections for which $q({\bm \theta_s}) > 2.71.$ We additionally apply the Asimov procedure to the null hypothesis to compute the expected limits~\cite{Cowan:2010js}, and power-constrain the reported limits~\cite{Cowan:2011an}, although the latter was not necessary in practice. We do not find any evidence of DM-induced He flash events in this analysis.

\begin{figure}[!t]
\includegraphics[width = 0.49\textwidth]{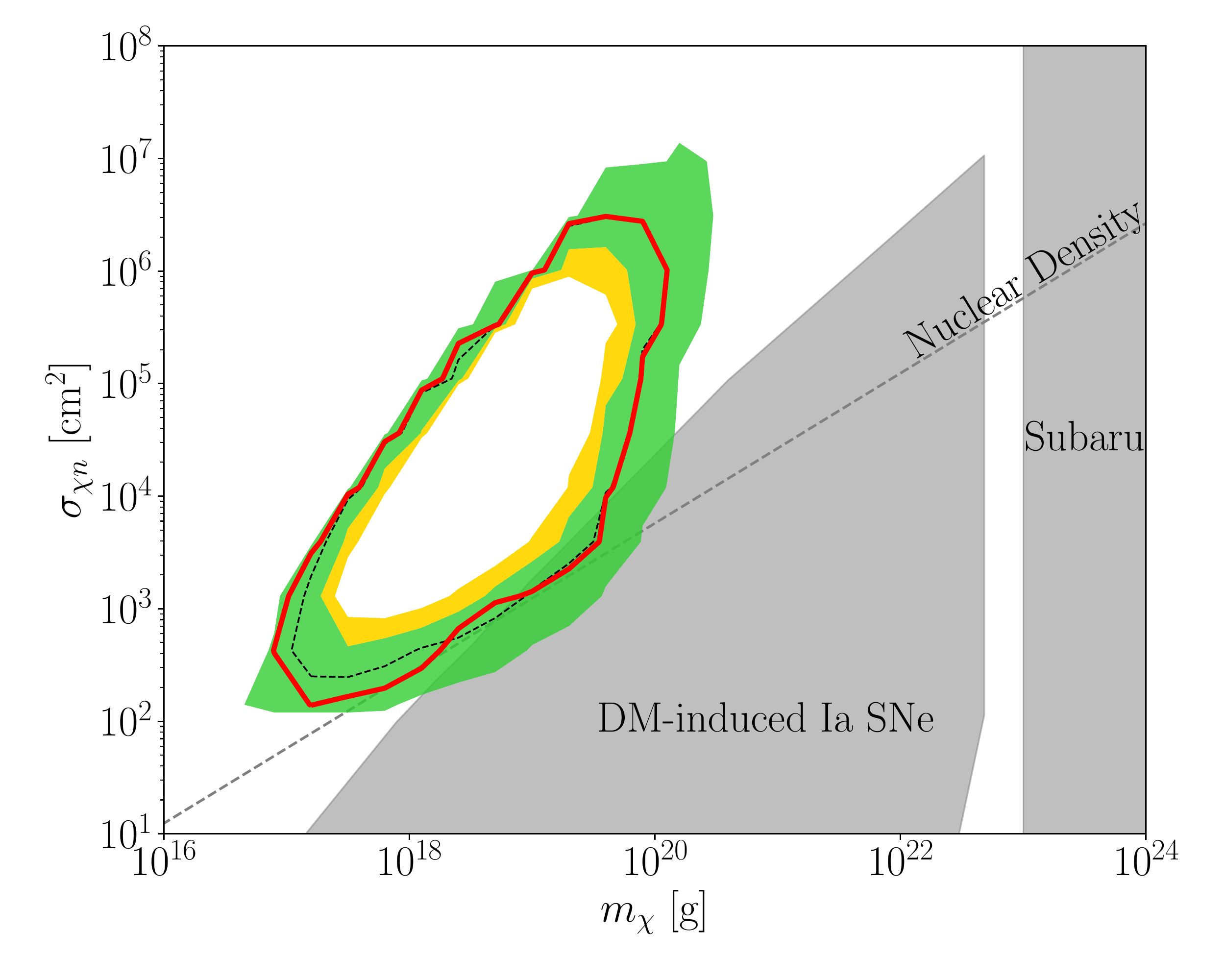}
\caption{ The red line is the 95\% limit on macroscopic DM from the non-observation of DM-induced He flashes in the GC M15 LF; the region inside this line is excluded. The dashed line is the Asimov expectation with green (yellow) bands denoting the 1$\sigma$ (2$\sigma$) containment region. We assume that the M15 DM is dominantly from the MW halo. We additionally show constraints from the non-observation of DM-induced Type Ia supernova in WDs~\cite{Janish:2019nkk,Sidhu:2019kpd} and from microlensing~\cite{Smyth:2019whb}. Macroscopic DM is bounded from above from CMB observations~\cite{Dvorkin:2013cea} and the non-observations of gas cloud heating~\cite{Bhoonah:2020dzs}, although at too large cross-section to be shown on this plot. We also show as a dotted line where the DM is nuclear density $\rho_0 = 2 \times 10^{14}$ g/cm$^3$.}
\label{fig:FinalLimit}
\end{figure} 

We show the 95\% limit on macroscopic DM from this search in Fig.~\ref{fig:FinalLimit}, and compare with the expected sensitivity under the null hypothesis from the Asimov procedure. The analogous plot assuming a GC overdensity is shown in Appendix Fig.~\ref{fig:FinalLimitGC}. At fixed $m_\chi$, the limits cut off at small $\sigma_{\chi n}$ below which the DM deposits too little energy to ignite helium fusion in the core. As $\sigma_{\chi n}$ increases, the signal increases until the cross section is so large that the DM is slowed by the stellar envelope. In that case, when the DM reaches the stellar core, it has too little kinetic energy remaining to ignite fusion, which cuts off the bounded region at large $\sigma_{\chi n}$. At high masses, the sensitivity decreases because the DM number density is too low for collisions to occur in the lifetime of the RGB stars. Nominally the above arguments suggest that the shape of the excluded region should be a right triangle, as discussed in Sec.~\ref{sec:rate}.  However, the time-dependence of the star over the RGB lifetime distorts this simple picture. At lower DM masses, the DM flux increases, so that we have increased sensitivity to DM-induced ignition in older stars. These stars can be ignited by DM with lower cross sections than their younger counterparts. Therefore, the bottom of the excluded region is slanted upwards rather than flat like the DM-induced ignition rates as in  Fig.~\ref{fig:Ignition Ratios}.

\section{Discussion}\label{sec:discussion}
We have outlined a process wherein macroscopic DM can collide with an RGB star core, depositing sufficient energy to ignite helium fusion. The result is the premature end to the RGB stage of stellar evolution for that star. The RGB luminosity function is a sensitive probe of this process; we use the LF of the GC M15 to constrain macroscopic DM properties.

A crucial assumption in our work is that the the DM-baryon scattering is purely elastic with geometric cross section. The elastic assumption breaks down if the binding energy of DM constituents is of order of the typical energy transfer $\sim$ MeV. This would unbind the DM, resulting in a smaller penetration depth, but increased local energy deposition as the DM breaks up. Even in the elastic scattering regime, whether or not the cross section is geometric depends on the DM substructure, the resulting form factor and the mean free path within the DM. It is left for future work to see how relaxing these assumptions could modify our constraints. For example, primordial black holes clearly do not obey this assumption \cite{Montero-Camacho:2019jte}, though we do not constrain black hole densities in any case. 

There are GC targets with improved sensitivity to this mechanism, although without existing LF data as in the case of M15. If a GC is found to host a DM halo, that GC would likely provide the strongest bound on macroscopic DM from this mechanism (see App.~\ref{app:GChalo} for an application of this scenario to M15). But we need not rely on the possibility of GC DM substructure. In particular, the GC M54 has a much larger DM density than that of M15 because it is located at the center of the Sagittarius dwarf galaxy. Ref.~\cite{Safdi:2018oeu} estimated that the Sagittarius DM density at the location of M54 is $\sim$7 GeV cm$^{-3}$, although if the DM profile is cusped the density could be orders of magnitude larger. Future measurements of the LF in M54 would have improved discovery potential over that of M15, in particular at high DM masses $m_\chi > 10^{21}$ g. The star clusters in the Galactic Center may also be sensitive probes of DM-induced HFs, given the extreme DM densities reached there. In fact, there is a lack of luminous RGB stars in the central $\sim$0.3 pc relative to expectations~\cite{1990ApJ...359..112S,2018A&A...609A..26G}, where there may be a DM spike around Sagittarius A~\cite{Merritt:2006mt}. Qualitatively, the missing red giants may be explained by DM-induced helium ignition, although we leave a quantitative assessment to future work and note that more standard explanations have been suggested~\cite{Bogdanovic:2013aka,2020MNRAS.492..250A,Dale:2008ff,2012ApJ...750..111A,Merritt:2005yt,Kim:2003hc,Zajacek:2020fgb}.

In this work we elected not to use the well studied tip of the red-giant branch to constrain the mechanism. The LF is more sensitive in the faint-signal regime where only a small percentage of high-luminosity RGB stars have been struck by DM; the slope of the LF can significantly increase even though the TRGB may be similar to that in the null case, for example if some RGB stars survive to undergo the standard HF. However, there can be some cases where the TRGB is affected, which would affect the calibration of the Hubble constant so that its value is dependent on the DM density in the vicinity of the RGB population.  

In future work, we will investigate additional probes of this mechanism. For instance, the early ignition of the HF leads to HB stars with smaller helium cores than typical of HB stars. Therefore, there may be a set of sub-HB stars at lower luminosity than the normal HB, consisting of stars which did not ascend the full RGB due to DM impacts. Additionally, the DM impacts may simultaneously ignite fusion in the whole core at once, whereas in the standard scenario the HF ignites in a series of layers over a longer timescale, which may affect direct observations of individual stars. However, observations have not yet resolved this transitionary period and distinguishing such stars from RGB or HB stars would likely prove difficult. 

In principle, one can also study a HF triggered by the accumulation and annihilation of DM. However, this mechanism is suppressed relative the analogous bound in WDs~\cite{Graham:2018efk} for two reasons. Firstly, RGB stars have a much lower DM capture rate than WDs due to their low core densities. Secondly, pure helium matter ignition requires a larger energy injection than the ignition of the carbon/oxygen matter of WDs. Then the annihilating DM must be more massive than in the WD case, so the DM number density is also suppressed.

\begin{acknowledgments}
{\it 
We thank Fred Adams for collaboration in the early stages of this work. We thank Holger Baumgardt for sharing data related to the M15 orbit. We thank Joshua Foster, Andrew Long, Aaron Pierce, and Benjamin Safdi for useful conversations. Z.J. was supported in part by DOE grant DE-SC0007859. C.D. was supported  in  part  by  the  DOE  Early Career  Grant  DE-SC0019225. C.D. and Z.J. were partially supported by the Leinweber Graduate Fellowship at the University of Michigan, Ann Arbor. This research was supported in part through computational resources and services provided by Advanced Research Computing at the University of Michigan, Ann Arbor.
}

\end{acknowledgments}

\bibliography{mybib}{}
\bibliographystyle{bibstyle}

\clearpage
\newpage 
\appendix
\renewcommand\thefigure{\thesection.\arabic{figure}}    
\setcounter{figure}{0}    
\section{M15 DM Halo}
\label{app:GChalo}
\begin{figure}[!t]
\includegraphics[width = 0.49\textwidth]{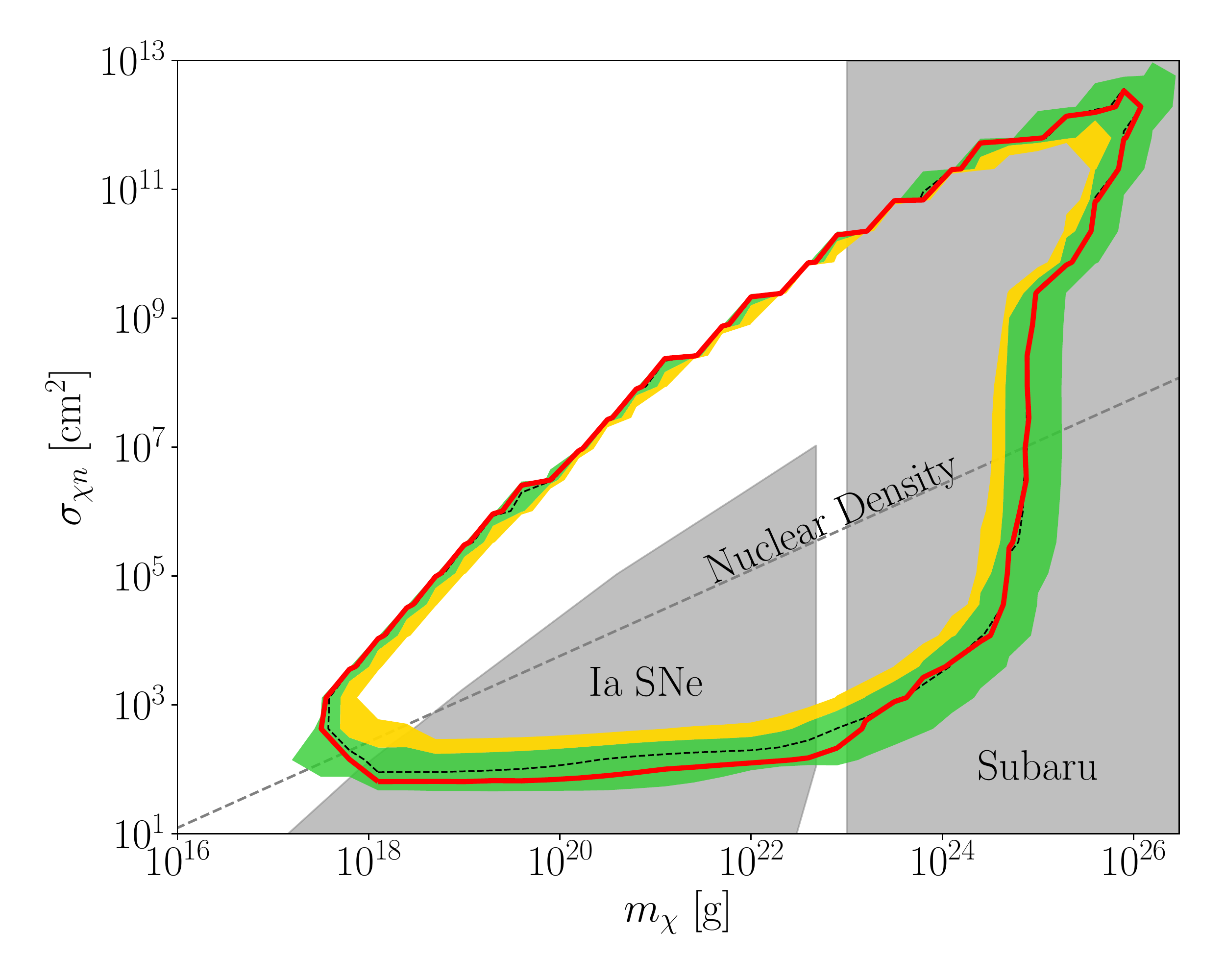}
\caption{The 95\% confidence limit on macroscopic DM (red) from the non-observation of DM-induced He flashes in the GC M15 LF, assuming that M15 hosts its own DM halo. The region inside the red line is ruled out. The dashed line is the Asimov expectation under the null hypothesis with green (yellow) bands denoting the 1$\sigma$ (2$\sigma$) containment region. We show the same constraints detailed in Fig.~\ref{fig:FinalLimit}.}
\label{fig:FinalLimitGC}
\end{figure}
In the main text we made the conservative assumption that the DM in M15 is sourced by the Milky Way halo. The DM density within GCs can be significantly larger if the GC formation took place in an initial DM overdensity, although this scenario is uncertain and we ignored it in our fiducial analysis. 
However, if further observations show there is indeed substructure, the constraints from DM-induced ignition can be improved, and we estimate the sensitivity gain in this section.

The formation of the GC occurs when the proto-GC virial temperature drops below $\sim 10^4$ K when radiative cooling turns on~\cite{2010MNRAS.405..375G}. At this time the proto-GC collapses, and the DM halo undergoes adiabatic contraction that increases the steepness of the density profile towards the center of the GC~\cite{1986ApJ...301...27B}. Simulations suggest that at high redshifts the subhalo around GCs is tidally stripped, such that the GC is baryon-dominated by mass at present times~\cite{Saitoh:2005tt}. At later times, stars kinetically heat the DM so that the profile becomes cored~\cite{Merritt:2003qk}.  

We assume the parameters derived in~\cite{HESS:2011yps}, which computed present-day M15 DM profile under this scenario in the context of DM annihilation searches. The DM distribution is cored at $\sim10$ pc \cite{HESS:2011yps}, inside which nearly all of the RGB stars are contained~\cite{1997ApJ...481..267D}.  Therefore, we take the DM density to be a constant throughout the core, with a value of
\begin{equation}
    \rho_{DM}= 35 M_\odot\ \rm{pc}^{-3} =  1.3 \times 10^3 \ \rm{GeV}\ \rm{cm}^{-3}.
\end{equation}

The radial stellar velocity dispersion $\sigma_{v_r}$ in M15 was measured within the inner $3$ pc by~\cite{1997ApJ...481..267D}; within the inner $1$ pc that reference found $\sigma_{v_r}$ to be a constant $10.2 \pm 1.4$ km s$^{-1}$, and declines to $\sim$7 km s$^{-1}$ at $3$ pc. In~\cite{Gerssen:2002iq} $\sigma_{v_r}$ was determined out to the cluster edge, where it continues declining to $\sim$5 km/s.  In the central $\sim$0.3 pc, $\sigma_{v_r}$ may rise to $\sim$14 km s$^{-1}$~\cite{Gerssen:2002iq}, but the number of stars in this region is volume suppressed.  Therefore, we adopt the conservative value $10$ km s$^{-1}$ as both the stellar and DM velocity dispersion, so that the relative velocity dispersion is $\sqrt{2}\times 10$ km s$^{-1}$. Lastly, the GC escape velocity of stars in this region is $\sim 50 $ km s$^{-1}$~\cite{HESS:2011yps}. Therefore it is a very good approximation to use Eq.~(\ref{eqn:MWrate}), setting $c(t)=1$, which gives
\begin{align}
 \Gamma &= \sqrt{2 \pi} \frac{\rho_{DM}}{m_{\chi}} R^2_{\zeta=1} \frac{ v^2_{*\text{esc}}(R_{\zeta=1}) }{ \sigma_{vr}}.
\end{align}
In Fig.~\ref{fig:FinalLimitGC}, we show the 95\% limit on macroscopic DM assuming that M15 hosts a DM subhalo with the above parameters. Similarly to our fiducial bound plot, the bounded region is roughly triangular, see text around Fig.\ref{fig:FinalLimit}. The extra tip to the triangle is due to the presence of friction near the hypotenuse, which for small values actually focuses the DM towards the core. Under this scenario, the constraints improve substantially, particularly at high $m_\chi$ where the rate is suppressed by low DM number density in the case where the DM is provided only by the Milky Way halo. The bound is nearly saturated in the sense that almost all of the parameter space in which DM-induced HFs are possible is disfavored, as $m_\chi > 10^{23}$ g is already disallowed by microlensing. To further sensitivity substantially, new targets hosting RGB stars of other masses or metallicities may be required, or targets with more RGB stars that are nearing the standard HF.

\section{Ignition Profile}
\setcounter{figure}{0}    
\label{app:ignition}
We assumed the energy is deposited in a linearly increasing profile from the edge of the geometric cross section to the center. The exact profile depends on the form of the DM, as well as on details of shock layer formation in front of the DM. This could increase the effective area over which the temperature is spread. In the limit where the only change is in the increase of the effective area, our bounds simply translate upwards to the correspondingly larger cross sections. 
 
 Separately one may be concerned of a sharp dependence of the ignition ratio, Eq.~\ref{eqn:IgnitionRatio}, on the shape of the profile. To investigate this we generalize Eq.~\ref{eqn:dEdV} in two ways. Defining $\xi=r/\sqrt{\sigma_{\chi n}/\pi}$, our nominal energy deposition is $\propto (1-\xi)$. We investigate the two sets of profiles
 \begin{align}
 \frac{dE}{dV}&\propto (1-\xi^n)\frac{dE}{dx}, \\
              &\propto (1-\xi)^n\frac{dE}{dx}.
 \end{align}
 \begin{figure*}[!t]
    \centering
     \includegraphics[width=1\textwidth]{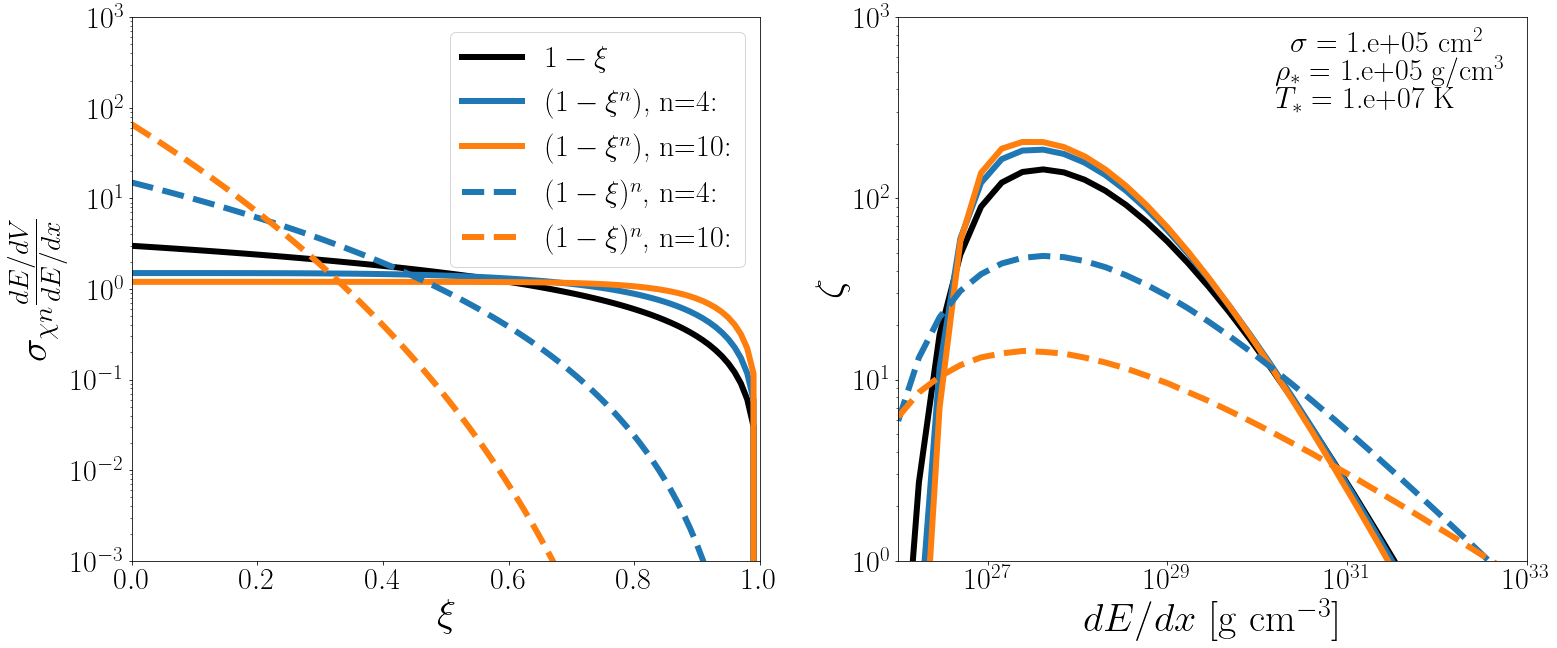}
     \caption{\textbf{Left Panel} The dimensionless energy deposition profiles. Each is normalized over a cylinder to give the same $dE/dx$.
     \textbf{Right Panel} The corresponding ignition parameter for various profiles. Stellar and DM properties are fixed at the values shown in the figure inset.}
     \label{fig:various_profiles}
 \end{figure*}
 In Fig.~\ref{fig:various_profiles} we show these distributions in the left panel, and an example of the resulting set of ignition ratios in the right panel. For $n=1$ these two shapes are equivalent, and is the case we assumed for our limits. In the large $n$ limit, the first set approaches a top hat shape, and the latter approaches a delta function peaked at the center of the DM ($\xi=0$). Assuming the former set of lines represent more physical DM models, we find very little shape dependence. 

\section{Analytic Ignition Rate}\label{app:rates}
\setcounter{figure}{0}    

\begin{figure}[!t]
    \centering
     \includegraphics[width=0.46\textwidth]{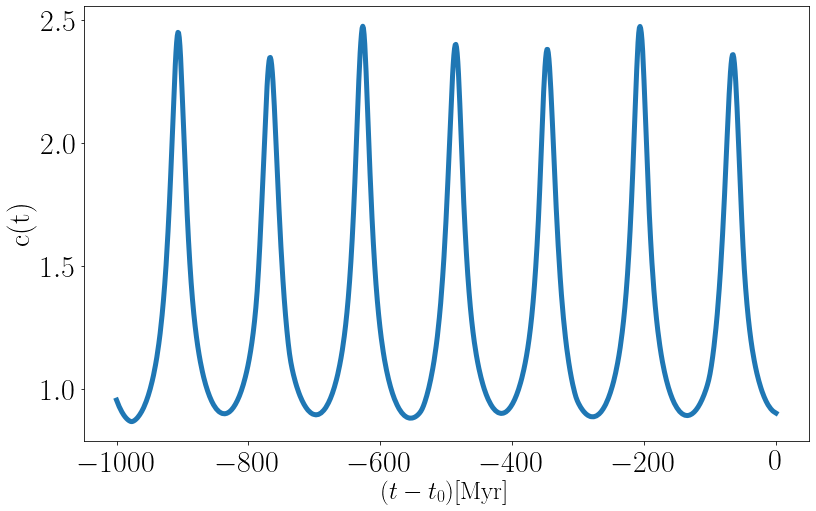}
     \caption{The correction factor $c(t)$ to the rate of DM-induced ignition for M15, see Eq.~\eqref{eqn:rate_details}.}
     \label{fig:correction}
 \end{figure}

In terms of the velocity distribution function, the rate of DM-induced ignition from an asymptotic surface $\Omega\ r_\infty^2$ is
\begin{align}
    \Gamma &= N  \frac{\rho_\chi}{m_\chi} r_{\infty}^2 \int d\Omega \int d^3{\bf v}\  (-\vec{v}\cdot \hat{r})\\ &\times \text{exp}\left(  \frac{ -({\bf v + v_{GC}})^2}{2 \sigma_{vr}^2} \right)\Theta( v^2_{esc}-({\bf v + v_{GC}})^2).
\end{align}
Where $\hat{r}$ points out from the star, and $N$ is the normalization of $f(\vec{v})$ 
\begin{align}
    N^{-1}&= (2 \pi \sigma_{vr}^2)^{3/2} \times k, 
\end{align}
For constant $k$ which accounts for finite MW escape velocity, given by 
\begin{align}
k= \left(  erf \left( \frac{v_{esc}}{\sqrt{2} \sigma_{vr}} \right) - \sqrt{\frac{2}{\pi}}\frac{v_{esc}}{\sigma_{vr}} \text{exp}\left(  \frac{ -v_{esc}^2}{2 \sigma_{vr}^2} \right)\right).
\end{align}
The solid angle of velocity space which hits the star is $\pi \phi^2 \approx \pi b^2/r_{\infty}^2 $ for impact parameter $b$. Conserving energy and angular momentum gives the gravitationally focused impact parameter as a function of the escape velocity from the core of the star $v_{*esc}$,
\begin{align}
    b^2=R^2_{\zeta=1}\left( 1+ v^2_{*esc}/v^2 \right)\approx R_{\zeta=1}^2 v^2_{*esc}/v^2.
\end{align}
With ${\bf v_{GC}}$ along the $z$-axis, ${\bf v \cdot v_{GC}}= v v_{GC} \text{ cos}\theta$. Applying these gives
\begin{align}
    \Gamma &= N  \frac{\rho_\chi}{m_\chi} (\pi R^2_{\zeta=1})v^2_{\text{esc}} 2 \pi \int^1_{-1} d \text{cos}\theta  \int^\infty_0 dv \ v \\ 
    &\times \text{exp}\left(  \frac{ -(v^2 + v_{GC}^2 + 2 v v_{GC} \text{cos}\theta )}{2 \sigma_{vr}^2} \right)\\ 
    &\times\Theta( v^2_{esc}-(v^2 + v_{GC}^2 + 2 v v_{GC} \text{cos}\theta)). 
\end{align}
We evaluate this integral with the methods in \cite{LEWIN199687} and arrive with 
\begin{align}
\Gamma &= \Gamma_0  \times c(t,v_{GC},v_{\text{esc}}).
\end{align}
Where $\Gamma_0$ is the rate including gravitational focusing, but without MW escape velocity, or GC velocity, evaluated at the current time $t_0$. The correction factor for the latter three effects defines $c(t,v_{GC}(t),v_{esc})$. These are
\begin{align}
\Gamma_0(\tau) &= \sqrt{2 \pi} \frac{\rho_{DM}(t_0)}{m_{\chi}} R^2_{\zeta=1}(\tau) \frac{v_{*esc}^2}{\sigma_{vr}} + \mathcal{O}(\sigma_{vr}^2/v^2_{*\text{esc}})\\
c(t)&= \frac{\rho_{DM} (t)}{\rho_{DM}(t_0)}\frac{\sqrt{\pi}\sigma_{vr}}{4 k v_{GC}} 
\Big [ 2 \sqrt{2} {\rm erf}(\frac{v_{GC}}{\sqrt{2}\sigma_{vr}}),  \label{eqn:rate_details}\\
&- {\rm Exp}\left(\frac{v_{esc}^2-2 v_{GC}^2}{2 \sigma^2_{vr}}\right) \nonumber\\
& \times \bigg({\rm erf}\left(\frac{v_{GC}+v_{esc}}{\sigma_{vr}}\right) - {\rm erf}\left(\frac{v_{esc}-v_{GC}}{\sigma_{vr}}\right) \bigg)   \Big ].\nonumber 
\end{align}
Using the M15 orbital data obtained from the authors of~\cite{Baumgardt2019}, we compute the correction factor $c(t)$ for M15 over the last Gyr and show it in Fig.~\ref{fig:correction}. This factor accounts for the DM density variation, GC velocity, and MW escape velocity as M15 orbits the Galactic Center. Minima (maxima) correspond to the orbital apoapses (periapses). At time $t_0$, today, the GC is near its apoapsis. We see that the rate is enhanced by a factor of $\sim$2.8 at periapsis relative to apoapsis.

\end{document}